\documentclass[final]{aipproc}
\usepackage{graphicx}
\usepackage[rflt]{floatflt}
\usepackage{mathptm}
\layoutstyle{6x9}

\renewcommand{\theequation}{\arabic{equation}}

\newcommand{\pvalt}{\raise0.15ex\hbox{-}\mkern-11.5mu\int}
\newcommand{\be}{\begin{equation}}
\newcommand{\ee}{\end{equation}}
\newcommand{\bea}{\begin{eqnarray}}
\newcommand{\eea}{\end{eqnarray}}
\newcommand{\ben}{\begin{enumerate}}
\newcommand{\een}{\end{enumerate}}
\newcommand{\bit}{\begin{itemize}}
\newcommand{\eit}{\end{itemize}}

\newcommand{\half}{\frac{1}{2}}

\newcommand\mn{{\mu\nu}}

\newcommand{\beq}{\begin{equation}}
\newcommand{\eeq}{\end{equation}}
\newcommand{\ba}{\begin{array}}
\newcommand{\ea}{\end{array}}

\begin{document}


\title{The Casimir Effect and the Quantum Vacuum}
 
\author{R.~L.~Jaffe}{
address={Center for Theoretical Physics,\\
Laboratory for Nuclear Science and Department of Physics\\
Massachusetts Institute of Technology, \\Cambridge, Massachusetts 02139},
}
\rightline{MIT-CTP-3614}

\begin{abstract} 

In discussions of the cosmological constant, the Casimir effect is often invoked as decisive evidence that the zero point energies of quantum fields are ``real''.  On the contrary, Casimir effects can be formulated and Casimir forces can be computed without reference to zero point energies.  They are relativistic, quantum forces between charges and currents.  The Casimir force (per unit area) between parallel plates vanishes as $\alpha$, the fine structure constant, goes to zero, and the standard result, which appears to be independent of $\alpha$, corresponds to the $\alpha\to\infty$ limit.


\end{abstract}

\maketitle 
 
\subsection{Introduction}
\label{section0} 
In  quantum field theory  as usually formulated, the zero point fluctuations of the fields contribute to the energy of the vacuum.  However this energy does not seem to be observable in any laboratory experiment.   Nevertheless, all energy gravitates, and therefore the energy density of the vacuum, or more precisely the vacuum value of the stress tensor, $\langle T_{\mu\nu}\rangle \equiv - {\cal E}g_{\mu\nu}$\footnote{I use the conventions of Ref.~\cite{weinbergbook} and in particular  take $g_{\mu\nu}={\rm diag}\ \{-1,1,1,1\}$, so ${\cal E}>0$ corresponds to a positive energy density and a negative pressure.}, appears on the right hand side of Einstein's equations,
\begin{equation}
\label{eq1}
R_{\mu\nu}-\frac{1}{2}g_{\mu\nu}R=-8\pi G(\tilde{T}_{\mu\nu}-{\cal E}g_{\mu\nu})
\end{equation}
where it affects cosmology.  ($\tilde T_{\mn}$ is the contribution of excitations above the vacuum.)  It is equivalent to adding a cosmological term, $ \lambda  =8\pi G{\cal E}$, on the left hand side. 

A small, positive cosmological term is now required to account for the observation that the expansion of the Universe is accelerating.  Recent measurements give\cite{Tegmark:2003ud}
\begin{equation}
\label{eq2}
\lambda = (2.14\pm 0.13 \times 10^{-3}\ \hbox{eV})^{4}
\end{equation}
at the present epoch.  This observation has renewed interest in the idea that the zero point fluctuations of quantum fields  contribute to the cosmological constant, $\lambda$\cite{zdv}.  However, estimates of the energy density due to zero point fluctuations exceed the measured value of $\lambda$ by many orders of magnitude.  Caution is appropriate when an effect, for which there is no direct experimental evidence, is the source of a huge discrepancy between theory and experiment!

As evidence of the ``reality'' of the quantum fluctuations of fields in the vacuum, theorists often point to the Casimir effect \cite{Casimir:1948dh}.  For example, Weinberg in his introduction to the cosmological constant problem, writes\cite{Weinberg:1988cp},
\begin{quote}
``Perhaps surprisingly, it was  along time before particle physicists began seriously to worry about [quantum zero point fluctuation contributions to $\lambda$] despite the demonstration in the Casimir effect of the reality of zero-point energies.''
\end{quote} 
More recent examples can be found in the widely read reviews by Carroll\cite{Carroll:2000fy},
\begin{quote}
`` ... And the vacuum fluctuations themselves are very real, as evidenced by the Casimir effect.''
\end{quote}
and by Sahni and Starobinsky \cite{Sahni:1999gb},\cite{further}
\begin{quote}
 ``The existence of zero-point vacuum fluctuations has been spectacularly demonstrated
by the Casimir effect.''
\end{quote} 

In 1997 Lamoreaux opened the door to precise measurement of Casimir forces\cite{Lamoreaux:1996wh}.  The Casimir force (per unit area) between parallel conducting plates,
\begin{equation}
\label{cas}
{\cal F} = -\frac{\hbar c\pi^{2}}{240d^{4}}
\end{equation}
has now been measured to about 1\% precision.
Casimir physics has become an active area of nanoscopic physics in its own right\cite{Bordag:2001qi}.  Not surprisingly, every review and text on the subject highlights the supposed special connection between the Casimir effect and the vacuum fluctuations of the electromagnetic field \cite{books}.

The object of this paper is to point out that the Casimir effect gives no more (or less) support for the ``reality'' of the vacuum energy of fluctuating quantum fields than any other one-loop effect in quantum electrodynamics, like the vacuum polarization contribution to the Lamb shift, for example.  The Casimir force can be calculated without reference to vacuum fluctuations, and like all other observable effects in QED, it vanishes as the fine structure constant, $\alpha$, goes to zero.

There is a long history and large literature surrounding the question whether the zero point fluctuations of quantized fields are ``real''\cite{milonni}.  Schwinger, in particular, attempted to formulate QED without reference to zero point fluctuations\cite{schwinger}.  In contrast Milonni has recently reformulated all of QED from the point of view of zero point fluctuations\cite{milonni}.  The question of whether zero point fluctuations of the vacuum are or are not ``real'' is beyond the scope of this paper.   Instead I address only the narrower question of whether the Casimir effect can be considered evidence in their favor.
\begin{floatingfigure}{.38\textwidth}
{\includegraphics[width=5cm]{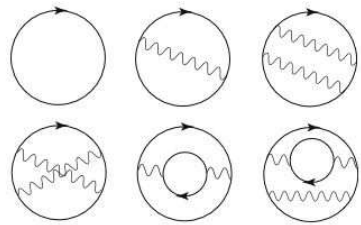}
\caption{\small QED graphs contributing to the zero point energy} \label{graphs}
}
\end{floatingfigure}

For a non-interacting quantum field the vacuum (or zero-point) energy is given by ${\cal E}=\pm\half\sum\hbar\omega_{0}$, where the $\{\hbar\omega_{0}\}$ are the eigenvalues of the free Hamiltonian and the plus or minus sign applies to bosons or fermions respectively.  In three dimensions the sum over frequencies diverges quartically, ${\cal E}\sim\Lambda^{4}$.  This contribution does not arise if the  fields in the classical Lagrangian are ordered in a prescribed way (``normal ordering''), but the reason for this choice of ordering is obscure and it is probably more appropriate to choose the ordering arbitrarily (though consistent with the symmetries of the theory), in which case the free field zero point energy can be cancelled by a counterterm.  However comparable contributions reappear when interactions are introduced:  the vacuum energy is related to the sum of all vacuum-to-vacuum Feynman diagrams, a few of which are shown ({\it eg.\/} for QED) in Fig.~\ref{graphs}.  A counter term can be introduced to cancel these contributions to any order in perturbation theory.  However since the leading divergence is quartic, such fine tuning is generally regarded as unacceptable.

In the standard approach\cite{books}, the Casimir force is calculated by computing the change in the zero point energy of the electromagnetic field when the separation between parallel perfectly conducting plates is changed.  The result, eq.~(\ref{cas}), seems universal, independent of everything except $\hbar$, $c$, and the separation, inviting one to regard it as a property of the vacuum.  This, however, is an illusion.  When the plates were idealized as perfect conductors, assumptions were made about the properties of the materials and the strength of the QED coupling $\alpha$, that obscure the fact that the Casimir force originates in the forces between charged particles in the metal plates.   More specifically,
\begin{itemize}
\item The Casimir effect is a function of the fine structure constant and vanishes as $\alpha \to 0$.  Explicit dependence on $\alpha$ is absent from eq.~(\ref{cas}) because it is an asymptotic form, exact in the $\alpha\to\infty$ limit.  The Casimir force is simply the (relativistic, retarded) van der Waals force between the  metal plates.  
\item Casimir effects in general can be calculated as $S$-matrix elements, {\it i.e.\/} in terms of Feynman diagrams with external lines, and without any reference to the vacuum or its fluctuations.  The usual calculation, based on the change in $\half\sum\hbar\omega$ with separation, is heuristic.  An elementary example of a similar situation occurs in electrostatics.  The energy of a smooth charge distribution, $\rho(x)$, can be calculated directly from $\half\int dxdy\frac{\rho(x)\rho(y)}{|x-y|}$, or alternatively, from the energy ``stored in the electric field'', $\frac{1}{8\pi}\int dx |\vec E(x)|^{2}$.  The existence of the latter formula cannot be regarded as evidence for the ``reality'' of the electric field, which awaited the discovery that light consists of propagating electromagnetic waves.
\end{itemize}

In the following section I review the dependence of the Casimir effect on the fine structure constant.  Next I discuss the calculation of Casimir effects without mention of vacuum energies.  Finally I conclude with a brief summary.   

\subsection{The dependence of the Casimir effect on the fine structure constant}

At first sight the Casimir force, eq.~(\ref{cas}) seems universal and independent of any particular interaction.  ${\cal F}$ depends only on the fundamental constants $\hbar$ and $c$.  However, a moment's thought reveals that interactions entered when one idealized the metallic plates as perfect conductors that impose boundary conditions on the electromagnetic fields.  

Actual metals are not perfect conductors.  In fact there is now a large literature dedicated to ``finite conductivity corrections'' to the Casimir effect\cite{books}. These treatments are based on Lifshitz' theory of the Casimir effect for dielectric media\cite{Lifshitz}.  A simpler treatment, based on the Drude model of metals, is sufficient describe things qualitatively\cite{am,jackson}.   A conductor is characterized by a  plasma frequency, $\omega_{\rm pl}$, and a skin depth, $\delta$.  $\omega_{\rm pl}$ characterizes the frequency above which the conductivity goes to zero. $\delta$ measures the distance that electromagnetic fields penetrate the metal.  Both $\omega_{\rm pl}$ and $\delta$ depend on the fine structure constant, $\alpha$, and vanish as $\alpha\to 0$.  In the Drude model,
\begin{eqnarray}
\label{eq4}
\omega_{\rm pl}^{2} &= &\frac{4\pi e^{2} n}{m}\nonumber\\
\delta^{-2} &=& \frac{2\pi \omega|\sigma|}{c^{2}}
\ \hbox{where}\ \sigma = \frac{ne^{2}}{m(\gamma_{0}-i\omega)}
\end{eqnarray}
where $n$ is the total number of conduction electrons per unit volume, $m$ is their effective mass, and $\gamma_{0}$ is the damping parameter for the Drude oscillators.  Typically the frequencies of interest are much greater than $\gamma_{0}$, so $\delta\approx c/\sqrt{2}\omega_{\rm pl}$.

The frequencies that dominate the Casimir force are of order $c/d$\cite{books}.  So the perfect conductor approximation is adequate if $c/d\ll \omega_{\rm pl}$, or 
\begin{equation}
\label{omegalimit}
\alpha\gg \frac{mc}{4\pi\hbar n d^{2}}.
\end{equation} 
Typical Casimir force measurements are made at separations of order 0.5 microns.  For a good conductor like copper, eq.~(\ref{omegalimit}) requires $\alpha$ to be greater than about $10^{-5}$, which is amply satisfied by the physical value $\alpha\approx 1/137$.  Thus the standard Casimir result can be regarded as the $\alpha\to\infty$ limit (!) of a result that for smaller values of $\alpha$ depends in detail on the nature of the plates.  

Let us examine the $\alpha\to 0$ limit.  In this limit the scale of atomic physics, the Bohr radius, $\hbar^{2}/me^{2}$, grows like $1/\alpha$.  Therefore $n$ scales like $e^{6}$ and both $\omega_{\rm pl}$ and $\delta$ vanish  like $\alpha^{2}$\footnote{Note, $\gamma_{0}\ll \omega$ remains a good approximation as $\alpha\to 0$ at fixed $d$.}.  So at any fixed separation, $d$, the Casimir force goes away quickly as $\alpha\to 0$.  Also, since $\delta\to\infty$ as $\alpha\to 0$, the separation, $d$, becomes ill-defined since the fields penetrate further than the nominal separation of the plates.  

The feature that distinguishes the Casimir force from many other effects in QED is that it reaches a finite limit as $\alpha\to \infty$.  Had that not been the case, the dependence on material parameters like $\omega_{\rm pl}$ would have had to be explicit and the effect would never have been accorded universal significance.  In fact just such a situation occurs in the case of the Casimir pressure on a conducting sphere.  If one calculates the Casimir pressure for a realistic material, one obtains a result that diverges as the plasma frequency (the cutoff on the $\omega$-integration) goes to infinity\cite{barton}.  Therefore it is impossible to define the Casimir pressure on a conducting sphere independent of the details of the material\footnote{This \emph{physical} problem must be distinguished from the \emph{mathematical} problem of the Casimir pressure on a perfectly conducting, perfectly spherical, zero thickness sphere considered by Boyer\cite{boyer,us}, which gives a finite result of no physical interest.}.

\newpage
\subsection{The Casimir effect without the vacuum}

Casimir's original goal was to compute the van der Waal's force between polarizable molecules at separations so large that relativistic (retardation) effects are essential.  He and Polder carried out this program and found an extremely simple result\cite{cp},
$$
\Delta E = -\frac{23\hbar c}{4\pi R^{7}}a_{1}a_{2}\ .
$$
\begin{floatingfigure}{.35\textwidth}
\vspace*{-.1in}
{\hspace*{.2in}\includegraphics[width=3cm]{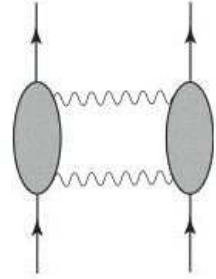}
\caption{\small Feynman diagrams for the Casimir-Polder force} \label{caspold}
}
\end{floatingfigure}  
\noindent 
 $a_{j}$ is the static polarizability of the $j^{\rm th}$ molecule, $\vec p=a \vec E$. They found a similarly simple result for a polarizable molecule opposite a conducting plate:  $\Delta E = -3\hbar ca/8\pi R^{4}$.  These results were derived using the standard apparatus of perturbation theory (to fourth order in $e$) without any reference to the vacuum.  They correspond to the long range limit of the Feynman diagrams of Fig.~\ref{caspold}.

Casimir was intrigued by the simplicity of the result, and following a suggestion by Bohr\cite{story}, showed that the Casimir-Polder results could be derived more simply by comparing the zero point energy of the electromagnetic field in the presence of the molecules with its vacuum values\cite{conf}.  He then considered the especially simple example where both molecules are replaced by conducting plates\cite{Casimir:1948dh}.  

Despite the simplicity of Casimir's derivation based on zero point energies, it is nevertheless possible to derive his result without any reference to zero point fluctuations or even to the vacuum.  Such a derivation was first given by Schwinger\cite{js1} for a scalar field, and then generalized to the electromagnetic case by Schwinger, DeRaad, and Milton\cite{sdm}.  Reviewing their derivation, one can see why the zero point fluctuation approach won out.  It is far simpler.  

In more modern language the Casimir energy can be expressed in terms of the trace of the Greens function for the fluctuating field in the background of interest ({\it e.g.\/} conducting plates),
\begin{equation}
\label{trace}
{\cal E} = \frac{\hbar}{2\pi}{\rm Im}\int d\omega 
\omega\ {\rm Tr}\int d^{3}x \left[{\cal G}(x,x,\omega+i\epsilon)-{\cal G}_{0}(x,x,\omega+i\epsilon)\right]
\end{equation}
where ${\cal G}$ is the full Greens function for the fluctuating field, ${\cal G}_{0}$ is the free Greens function, and the trace is over spin.  

On the one hand 
$$\frac{1}{\pi}{\rm Im}\int [{\cal G}(x,x,\omega+i\epsilon)-{\cal G}_{0} (x,x, \omega+ i\epsilon)]=\frac{d\Delta N}{d\omega}
$$
 is the change in the density of states due to the background, so eq.~(\ref{trace}) can be regarded as a restatement of the Casimir sum over shifts in zero-point energies, $\frac{1}{2} \sum (\hbar\omega-\hbar\omega_{0})$.  On the other hand, the Lippman-Schwinger equation allows the full Greens function, ${\cal G}$, to be expanded as a series in the free Green's function, ${\cal G}_{0}$, and the coupling to the external field as in Fig.~ \ref{greensfunction} \footnote{This reformulation is based on the identification of the Casimir energy as the one-loop \emph{effective action} in a static background, $\sigma$, ${\cal S[\sigma]}=T{\cal E}[\sigma]$\cite{Peskin,Weinbergqft}.}.  So the Casimir energy can be expressed entirely in terms of Feynman diagrams with external legs --- {\it i.e.\/} in terms of $S$-matrix elements which make no reference to the vacuum.  
\begin{floatingfigure}{.6\textwidth}
{\includegraphics[width=8cm]{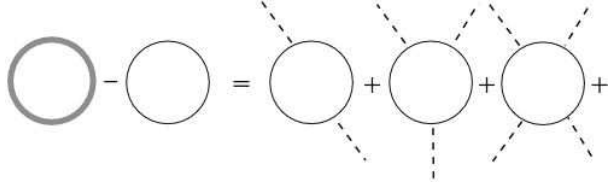}
\caption{\small Diagrammatic expansion of the Casimir force: The thick (thin) line denotes the full (free) Greens function; the one-point function is omitted because it does not contribute to the force. } \label{greensfunction}
}
\end{floatingfigure}   

As an explicit example, consider the Casimir effect for a  scalar field, $\phi$, in one dimension, forced to obey a Dirichlet boundary condition, $\phi=0$, at $x=\pm a/2$.  A traditional calculation, summing over zero-point energies, yields a Casimir force, $F=-\hbar c\pi/24a^{2}$ in this case.  This is the one-dimensional, scalar analogue of Casimir's original calculation.  As in that case, the dependence on the coupling constant has been obliterated by taking a limit where the interaction can be idealized as a boundary condition.  To better model the physical situation we replace the boundary condition by a $\delta$-function external potential at $x=\pm a/2$.  Explicitly, we calculate the force between the singular points at $x=\pm a$ by calculating the derivative with respect to $a$ of the effective energy of $\phi$ in the presence of a background field, $\sigma(x)$.  The interaction is 
\begin{equation}
\label{phisigma}
{\cal L}_{\rm int} = \half g\sigma(x)\phi^{2}(x)
\end{equation}
and we specify $\sigma(x)=\delta(x-a/2)+\delta(x+a/2)$.  The ``boundary condition limit'', $\phi(\pm a/2)=0$, is obtained by sending $g\to\infty$.  To regulate infrared divergences that afflict scalar fields in one dimension, we introduce a mass, $m$, for $\phi$.

The effective energy is given by the sum of all one loop Feynman diagrams with insertions of $\sigma(x)$ --- the diagrams shown in Fig.~\ref{greensfunction} --- and its derivative with respect to $a$ gives the force\cite{books,us},
\begin{equation}
\label{onedim}
F(a,g,m)=-\frac{g^{2}}{ \pi}
\int_m^\infty\frac{t^{2}dt}{\sqrt{t^{2}-m^{2}}} \ 
\frac{e^{-2at}}{4t^{2}+4g t +g^{2}(1-e^{-2at})
} 
\end{equation}
This result embodies all the features we desire.  It vanishes (quadratically) as $g \to 0$, as expected for a phenomenon  generated by the coupling of $\phi$ to the external field.  In the boundary condition limit, $g\to\infty$, the dependence on the material disappears,
\begin{equation}
\label{dir}
\lim_{g\to\infty}F(a,g,m)  
=-\int_m^\infty \frac{dt}{\pi}
\frac{t^2}{\sqrt{t^2-m^2} \left(e^{2ta}-1\right)}\, ,
\end{equation}
and it reduces to $- \pi/24 a^{2}$ in the $m\to 0$ limit.  
\newpage
\subsection{Conclusion}

I have presented an argument that the experimental confirmation of the Casimir effect does not establish the reality of zero point fluctuations.  Casimir forces can be calculated without reference to the vacuum and, like any other dynamical effect in QED,  vanish as $\alpha\to 0$.  The vacuum-to-vacuum graphs (See Fig.~\ref{graphs}) that define the zero point energy do not enter the calculation of the Casimir force, which instead only involves graphs with external lines. So the concept of zero point fluctuations is a heuristic and calculational aid in the description of the Casimir effect, but not a necessity.    

The deeper question remains:  Do the zero point energies of quantum fields contribute to the energy density of the vacuum and, \emph{mutatis mutandis}, to the cosmological constant?  Certainly there is no experimental evidence for the ``reality'' of zero point energies in quantum field theory (without gravity).   Perhaps there is a consistent formulation of relativistic quantum mechanics in which zero point energies never appear.  I doubt it.  Schwinger intended source theory to provide such a formulation.  However, to my knowledge no one has shown that source theory or another $S$-matrix based approach can provide a complete description of QED to all orders.  In QCD confinement would seem to present   an insuperable challenge to an $S$-matrix based approach,  since quarks and gluons do not appear in the physical $S$-matrix.  Even if one could argue away quantum zero point contributions to the vacuum energy, the problem of spontaneous symmetry breaking remains:  condensates that carry energy appear at many energy scales in the Standard Model.  So there is good reason to be skeptical of attempts to avoid the standard formulation of quantum field theory and the  zero point energies it brings with it.  Still,  no known phenomenon, including the Casimir effect, demonstrates that zero point energies are ``real''.

\subsection{Acknowledgments}

I would like to thank Eddie Farhi, Jeffrey Goldstone, Roman Jackiw, Antonello Scardicchio, and Max Tegmark for conversations and suggestions, and the Rockefeller Foundation for a residency at the Bellagio Study and Conference Center on Lake Come, Italy, where this work was begun.  This work is  
supported in part by funds provided by the U.S.~Department of
Energy (D.O.E.) under cooperative research agreement
DE-FC02-94ER40818.



  \end{document}